\documentclass[12pt]{article}
\usepackage{graphicx}
\usepackage{amsbsy}
\usepackage[all]{xy}
\usepackage{amsmath}
\makeatletter
\newcommand{\singlespacing}{\let\CS=\@currsize\renewcommand{\baselinestretch}
{1.0}\tiny\CS}
\newcommand{\doublespacing}{\let\CS=\@currsize\renewcommand{\baselinestretch}
{1.5}\tiny\CS}

\newcommand{\beq}{\begin{equation}}
\newcommand{\eeq}{\end{equation}}

\begin{document}

\begin{center}
{\bf \Large Spin Transport in non-inertial frame}
\end{center}

\begin{center}

Debashree Chowdhury  \footnote{Corresponding author, Tel.: +91 3325753037; fax: +91 3325753026.}$^{,~2}$ and B Basu
 \footnote { E-mail:~~{debashreephys@gmail.com (D. Chowdhury)} and  sribbasu@gmail.com(B. Basu)} \\
 Physics and Applied Mathematics Unit\\ \
Indian Statistical Institute, Kolkata \\ \
203, B. T Road, Kolkata - 700108, India\\ \


\end{center}
\begin{abstract}

The influence of acceleration and rotation on spintronic applications is theoretically investigated. In our formulation, considering a Dirac particle in a non-inertial frame, different spin related aspects are studied. The spin current appearing due to the inertial spin-orbit coupling (SOC) is enhanced by the interband mixing of the conduction and valence band states. Importantly, one can achieve a large spin current through the $\vec{k}. \vec{p}$ method in this non-inertial frame. Furthermore, apart from the inertial SOC term due to acceleration, for a particular choice of the rotation frequency, a new kind of SOC term can be obtained from the spin rotation coupling (SRC). This new kind of SOC is of Dresselhaus type and controllable through the rotation frequency. In the field of spintronic applications, utilizing the inertial SOC and SRC induced SOC term, theoretical proposals for the  inertial spin filter, inertial spin galvanic effect are demonstrated. Finally, one can tune the spin relaxation time in semiconductors by tuning the non-inertial parameters.

\end{abstract}


$keywords:$  inertial spin orbit coupling, spin current, spin filter, spin Galvanic effect, spin relaxation time

\section{Introduction}
$Spintronics$ or spin based electronics has received enormous attention in the last few years due to its several remarkable applications in various areas of physics, particularly to the condensed matter systems. In this broad area of research, one studies the quantum transport properties of the electron spins and its application to technology \cite{wolf,zutic,sh1}. The main challenge of the theory arises because of the fact that the spin current is a non-conserved quantity and the control and generation of spin current is a very difficult task. Since the theoretical prediction of the spin Hall effect (SHE) \cite{spinh}, which is a form of anomalous Hall effect induced by spin, the spin related physics has gained considerable advancement. SHE is the separation of the electrons, having up and down spin projections in the presence of a perpendicular electric field in analogy to Hall effect where charges are separated due to the application of a perpendicular magnetic field. Besides, the spin orbit interaction, which plays a crucial role in semiconductor spintronics, opens up the possibility of manipulating electron (or hole) spin in semiconductors by electrical means \cite{29, 30} and as such attracted a lot of interest of theoreticians and experimentalists recently.

However, in the arena of spin transport issues, except a few \cite{matsuoprb,cb,papini},the effect of the spin-orbit interaction (SOI) in non-inertial frames has not much been addressed in the literature, although studies on the inertial effect \cite{barnett,ein,tol,o} of electrons is not new.
Inclusion of the inertial effects in semiconductors can open up some fascinating phenomena, yet not addressed and hence   
it is appealing to investigate the role of these inertial effects on some aspects of spin transport in semiconductors. 

On the other hand, apart from the spin orbit coupling (SOC), there exists other two types of spin coupling as Zeeman coupling and spin rotation coupling (SRC)\cite{matsuoprl}, where spin of electron couples with other parameters. It is interesting to note that if in a non-inertial frame, the rotation frequency is considered to vary with momentum and time, then it is possible to generate a SRC induced SOC term.  This new type of SOC term is of Dresselhaus type \cite{dress} and controllable through mechanical rotation \cite{src}. Unlike the Rashba \cite{rashba} coupling, Dresselhaus coupling is not controllable through the external electric or magnetic fields. This can open up a new idea of controllable Dresselhaus like effect in semiconductor nano-structure at least theoretically.
 
Furthermore, the spin dynamics of the semiconductor is influenced by the $\vec{ k} . \vec{ p}$ perturbation theory.
On the basis of $ \vec{ k} . \vec{ p}$ perturbation theory, one can reveal many characteristic features related to spin dynamics.
We have theoretically demonstrated the generation of spin current in a solid on the basis of
$\vec{ k} . \vec{ p}$ perturbation \cite{winkler} with a generalized spin orbit Hamiltonian which includes the inertial effect due to acceleration. Considering the $\vec{ k} . \vec{p}$ perturbation in the $8 \times 8$ Kane model, the generation of spin current is studied in the extended Drude model framework, where the SOC has played an important role.
The interplay of Aharonov-Bohm phase ($AB$) and Aharonov-Casher ($AC$)  like phases helps us to propose a perfect spin filter for the accelerating system. Taking into account different $k$ linear inertial coupling terms, we have given a proposal for inertial spin galvanic effect. 

Besides, it is worth studying the variable spin relaxation times in spintronic devices. It is shown here how the spin relaxation time is affected in a non-inertial frame. 
 
The paper is organized as follows. 
Sec 2 contains the derivation of the non-relativistic Hamiltonian in presence of acceleration and rotation and the calculation of the spin current in presence of $\vec{k}.\vec{p}$ method. In Sec 3 we propose a spin filter configuration in our non-inertial system. In the next section (Sec 4) we have pointed out how the SRC induced SOC term can play a crucial role for the spin physics. Sec 5 deals with the inertial spin galvanic effect whereas in sec. 6 we discuss the effect of non-inertial parameters on spin relaxation time. Finally, we conclude in sec 7.

\section{Spin-orbit coupling and generation of spin current in non-inertial frames}
One of the main challenges in the study of spin based electronics lies with the fact that the spin current is 
a non-conserved quantity and the control and generation of spin current is a very difficult job. 
In a very recent paper \cite{papini}, it has interestingly been shown that the non-inertial fields can be 
used to generate and control spin currents.               
In this regard, one  can start with the covariant Dirac equation in the presence of acceleration and rotation as
\beq \left[i\gamma^{\mu}(x)(D_{\mu} + ieA_{\mu}) - m \right]\Psi(x) = 0,\eeq where $D_{\mu} = \partial_{\mu} + i\Gamma_{\mu}(x)$ is the covariant derivative and $\Gamma_{\mu}(x)$ are the spin connections containing the effect of rotation($\vec{\Omega}$) and acceleration($\vec{a}$) terms, $A_{\mu}$'s are the gauge fields and $m$ is the mass of the particle. 

The transfer of angular momentum between the external non-inertial fields and the electron spins are studied by the standard definition of the spin current tensor given by \cite{book}
 \beq S^{\rho\mu\nu} = \frac{1}{4im}\left[(\partial^{\rho}\overline{\Psi})\sigma^{\mu\nu(x)}\Psi - \overline{\Psi}\sigma^{\mu\nu(x)}(\partial^{\rho}\Psi)\right],\eeq
 where $\Psi$ is the spinor wave function and $\sigma^{\mu\nu}(x) = \frac{i}{2}[\gamma^{\mu},\gamma^{\nu}],$ where $\gamma^{\mu}= e^{\mu}_{~~\alpha}(x)\gamma^{\alpha}.$ Here $e^{\mu}_{~~\alpha}$ are the vierbeins and  $\gamma^{\alpha}$ are the usual Dirac matrices. It is possible to write $S^{\rho\mu\nu}$ as composed of two parts corresponding to inertial and non-inertial contributions \cite{papini}. In analogy to the external electromagnetic fields it has been shown that the external non-inertial fields invalidate the conservation of the spin current tensor.
 In the rest frame of the particle, with a particular choice of $\vec{\Omega} = (0,0,\Omega)$ and a detailed calculation shows\cite{papini}  
  \beq \partial_{\rho} S^{\rho\mu\nu} = \partial_{i}S^{i12} = 
  \frac{E+m}{2E}\frac{\Omega a_{2}x}{1+ \vec{a}.\vec{x}},\eeq  $\partial_{i}S^{i13}=\partial_{i}S^{i23}=0$ where $a_{2}$ is the
  acceleration in the $y$ direction, E and m are the energy and mass of the particle respectively. 
In the absence of acceleration or rotation or both the conservation of the spin current tensor is restored.

In a different approach, using the non-relativistic limit of the Dirac Hamiltonian in a non-inertial frame \cite{o}, where the inertial SOC plays a very crucial role, the generation of inertial spin current has been studied in \cite{cb}.

In this context, let us construct the Dirac Hamiltonian in
a non-inertial frame using the tetrad formalism, following the work of Hehl and Ni \cite{o} as
 \begin{eqnarray}\label{ha}
 H = \beta mc^{2} +  c\left(\vec{\alpha}.\vec{p}\right)
 +\frac{1}{2c}\left[(\vec{a}.\vec{r})(\vec{p}.\vec{\alpha}) +
 (\vec{p}.\vec{\alpha})(\vec{a}.\vec{r})\right]
 +\beta m
 (\vec{a}.\vec{r}) - \vec{\Omega}.(\vec{L} + \vec{S})\label{q}
 \end{eqnarray}
 where $\vec{a}$ and  $\vec{\Omega}$ are respectively the linear acceleration and
 rotation frequency of the observer with respect to an inertial frame.  $\vec{L}(= \vec{r}\times \vec{p}$)
  and $\vec S$ are respectively  the orbital angular momentum
  and spin of the Dirac particle. $ \beta$ and $\vec{\alpha} $ are the Dirac matrices.
The non-relativistic limit of the Hamiltonian is obtained by the application of a series of Foldy-Wouthuysen transformations (FWT) \cite{gre,m}.  The Pauli-Schroedinger Hamiltonian of a Dirac particle for positive energy solution in the inertial frame is given as
\begin{equation}\label{hb}
 H_{FW} = \left( mc^{2} + \frac{\vec{p}^{2}}{2m}\right) +  m (\vec{a}.\vec{r})
 +\frac{\hbar}{4mc^{2}}\vec{\sigma}.(\vec{a}\times \vec{p}) - \vec{\Omega}.(\vec{r}\times \vec{p} + \vec{S}),
\end{equation}
 where $m (\vec{a}.\vec{r})$ and $\vec{\Omega}.(\vec{r}\times \vec{p})$ are the inertial potentials due to acceleration and rotation respectively.
 Interestingly, the potential $V_{\vec{a}} = -\frac{m}{e}\vec{a}.\vec{r}$ can induce an electric field as $\vec{E}_{a}$ \cite{matsuoprb,cb,bcs, bc}, given by
 \beq \vec{E}_{a} = \frac{m}{e}\vec{a} = -\vec{\nabla} V_{a}(\vec{r}).\eeq  Similarly, the rotation of the frame of reference can induce a magnetic field  as
\beq\vec{B}_{\Omega} = \vec{\nabla}\times \vec{A}_{\Omega} = 2\frac{m}{e}\vec{\Omega},\eeq which has an analogous  form of the gravitomagnetic field, where
$\vec{A}_{\Omega} = \frac{mc}{e}(\vec{\Omega}\times\vec{r})$ is the gauge potential corresponding to the magnetic field induced by rotation.
The Hamiltonian (5), in terms of the induced electric and magnetic fields due to inertial effect, is as follows
 \beq H_{FW} = \frac{\vec{p}^2}{2m} - eV_{a}(\vec{r}) -\vec{\Omega}.(\vec{r}\times \vec{p}) - \frac{e\hbar}{4m^{2}c^{2}}\vec{\sigma}.\left(\vec{p} \times \vec{E}_{a}\right) - \frac{e\hbar}{4m}\vec{\sigma}.\vec{B}_{\Omega} ,\label{hc} \eeq where the third term in the r.h.s of Hamiltonian (\ref{hc}) is the coupling term between rotation and orbital angular momentum. The gauge potential $\vec{A}_{\Omega}$ originates from this coupling term. The forth term in the r.h.s of (\ref{hc}) is the inertial spin orbit coupling term. On the other hand, the Zeeman like coupling term (last term in the r.h.s of (\ref{hc}))
  is due to the spin rotation coupling. The acceleration induced
 electric field has important contribution on the spin current and spin polarization \cite{cb,bc}.
Retaining the terms upto $\frac{1}{c^{2}}$ order, the Hamiltonian in (\ref{hc}) can be rewritten as
 \beq H_{FW} = \frac{1}{2m}\left(\vec{p} - \frac{e}{c}\vec{A}_{\Omega} - \frac{\mu}{2c} \vec{E}_{a}\times\vec{\sigma}\right)^{2} - \phi_{I} -
 \frac{\mu}{2}\sigma . \vec{B}_{\Omega} ,\label{ab}\eeq where $\mu = \frac{e\hbar}{2mc}$ is the magnetic moment of electron and $\phi_{I} = e(V_{a} + \phi_{\Omega}),$ is the total inertial potential of the system where $\phi_{\Omega} = \frac{m}{2e}(\vec{\Omega}\times\vec{r})^{2}$ is the
 induced potential due to rotation.
 This Hamiltonian has $U(1)\otimes SU(2)$ gauge symmetry with $U(1)$ gauge potential $A_{\mu} = (\phi_{I} ,\vec{ A}_{\Omega})$ and $SU(2)$
 spin gauge potential $b_{\mu} = (-\vec{\sigma} .\frac{\vec{B}_{\Omega}}{2} , - \vec{\sigma}\times \frac{\vec{E}_{a}}{2}).$
 This $U(1)\otimes SU(2)$ gauge theory gives an unified picture of charge and spin dynamics.

Within the extended Drude model framework, from this FW transformed Hamiltonian we proceed to derive the expression of spin current by taking into account the $\vec{k} . \vec{p}$ theory. The fact that physical parameters in vacuum are renormalized when considered within a solid, inspires us to study the renormalization effects of inertial spin current in a crystalline solid in the framework of $\vec{k} . \vec{p}$ perturbation theory taking into account the Kane model \cite{kane} using the Bloch eigenstates.


In $ 8\times 8 $ Kane model the $\vec{ k} . \vec{ p}$ coupling between the $\Gamma_{6} $ conduction band and $\Gamma_{8} $ and $\Gamma_{7} $ valance bands is considered \cite{winkler}. Thus with the acceleration induced electric field and an external electric  $(\vec{E} = -\vec{\nabla} V(r))$ and magnetic field $(\vec{B})$ we can write the Hamiltonian as
\begin{eqnarray}
H_{8 \times 8}  =  \left( \begin{array}{ccr}
H_{6c6c} & H_{6c8v} & H_{6c7v} \\
H_{8v6c} & H_{8v8v} & H_{8v7v}\\
H_{7v6c} & H_{7v8v} & H_{7v7v}
\end{array} \right)~~~~ \\
~~=  \left( \begin{array}{ccr}
(E_{c} + eV_{tot})I_2 &\sqrt{3}P\vec{ T} . \vec{ k}& -\frac{P}{\sqrt{3}}\vec{ \sigma} . \vec{ k} \\
\sqrt{3}P \vec{ T}^{\dag} . \vec{ k}& (E_{v} + eV_{tot})I_4 & 0 \\
-\frac{P}{\sqrt{3}}\vec{ \sigma} . \vec{ k}& 0 & (E_{v} - \triangle_{0} + eV_{tot})I_{2}
\end{array} \right).
\end{eqnarray}
Here, $V_{tot} = V(\vec{r}) - V_{a}(\vec{r}),$ $E_{c}$ and $E_{v}$ are the energies at the conduction and valence band edges respectively. $ \triangle_{0},$ P are the spin orbit gap and the Kane momentum matrix element respectively. $ \triangle_{0}$ and $E_{G}=E_c-E_v$ varies with materials, whereas P is almost constant for group III to V semiconductors. The $\vec{T}$ matrices are given by
\begin{eqnarray}
T_{x}  &=& \frac{1}{3\sqrt{2}} \left( \begin{array}{ccrr}
-\sqrt{3} & 0 & 1 & 0 \\
0 & -1 & 0 & \sqrt{3}
\end{array} \right),\nonumber\\
T_{y}  &=& -\frac{i}{3\sqrt{2}} \left( \begin{array}{ccrr}
\sqrt{3} & 0 & 1 & 0 \\
0 & 1 & 0 & \sqrt{3}
\end{array} \right),\nonumber\\
T_{z}  &=& \frac{\sqrt{2}}{3} \left( \begin{array}{ccrr}
0 & 1 & 0 & 0 \\
0 & 0 & 1 & 0
 \end{array}\right)
\end{eqnarray}
and $I_{2}, I_{4}$ are the unit matrices of size $2$
and $4$ respectively.


In absence of the rotation term, the inertial FW Hamiltonian (8) can be reduced to an effective Hamiltonian of the conduction band electron states \cite{winkler} as
\begin{eqnarray}
H_{kp} &=& \frac{P^2}{3}\left(\frac{2}{E_{G}} + \frac{1}{E_{G} + \triangle_{0}}\right)\vec{k}^{2} + eV_{tot}(\vec{r})\nonumber\\
&& - \frac{P^2}{3}\left(\frac{1}{E_{G}} - \frac{1}{(E_{G} + \triangle_{0})}\right)\frac{ie}{\hbar}\vec{\sigma}.(\vec{k}\times \vec{k})\nonumber\\ &+ & e\frac{P^2}{3}\left(\frac{1}{E_{G}^{2}} - \frac{1}{(E_{G} + \triangle_{0})^{2}}\right)\vec{\sigma}.(\vec{k}\times \vec{E}_{tot})
\end{eqnarray}
The total inertial Hamiltonian i.e including the effect of acceleration for the conduction band electrons is then given by,
\begin{eqnarray}
 H_{tot} &=& \frac{\hbar^{2}\vec{k}^{2}}{2m^*} + eV_{tot}(\vec{r}) + (1 + \frac{\delta g}{2})\mu_{B}\vec{\sigma} . \vec{B} \nonumber\\
 &&+ e(\lambda + \delta \lambda)\vec{ \sigma} .(\vec{ k}\times \vec{ E}_{tot}) ,\label{Hkp }
\end{eqnarray}
 where
$\frac{1}{m^*} = \frac{1}{m} + \frac{2P^2}{3\hbar^{2}}\left(\frac{2}{E_{G}} + \frac{1}{E_{G} + \triangle_{0}}\right)$ is the effective mass and
$\vec{ E}_{tot} = -\vec{ \nabla} V_{tot}(\vec{r}) = \vec{E} - \vec{E}_{a},$ is the effective total electric field of the
inertial system and $ \lambda = \frac{\hbar^{2}}{4m^{2}c^{2}}$ is the SOC strength as considered in vacuum. The perturbation parameters $\delta g$ and $\delta \lambda$ are given by
\begin{eqnarray}\label{lam}
\delta g &=& -\frac{4m}{\hbar^{2}}\frac{P^2}{3}\left(\frac{1}{E_{G}} - \frac{1}{E_{G} + \triangle_{0}}\right)\nonumber\\
\delta \lambda &=& + \frac{P^2}{3}\left(\frac{1}{E_{G}^{2}} - \frac{1}{(E_{G} + \triangle_{0})^{2}}\right),
\end{eqnarray}
are the renormalized Zeeman coupling strength and renormalized SOC parameter respectively. These terms appear due to the interband mixing on the basis of ${\vec k}.{\vec p}$ perturbation theory. The total Hamiltonian have the following form
\beq H_{tot} = \frac{\hbar^{2}k^{2}}{2m^*} + eV_{tot} + (1 + \frac{\delta g}{2})\mu_{B}\vec{\sigma} . \vec{B} + e\lambda_{eff}\vec{\sigma} .(\vec{ k}\times \vec{ E}_{tot}) ,\label{eff} \eeq where $\lambda_{eff} = \lambda + \delta\lambda$ is the effective SOC term, which plays an important role in the rest part of our derivation. 
 In the absence of an external magnetic field the relevant part of the Hamiltonian can be written as
\begin{equation}\label{124}
H =  \frac{\vec{p}^{2}}{2m^*} + eV_{tot}(\vec{r})
- \lambda_{eff}\frac{e}{\hbar}\vec{\sigma}.(\vec{E}_{tot}\times \vec{p})
\end{equation}

\begin{figure}
\includegraphics[width=6.0 cm]{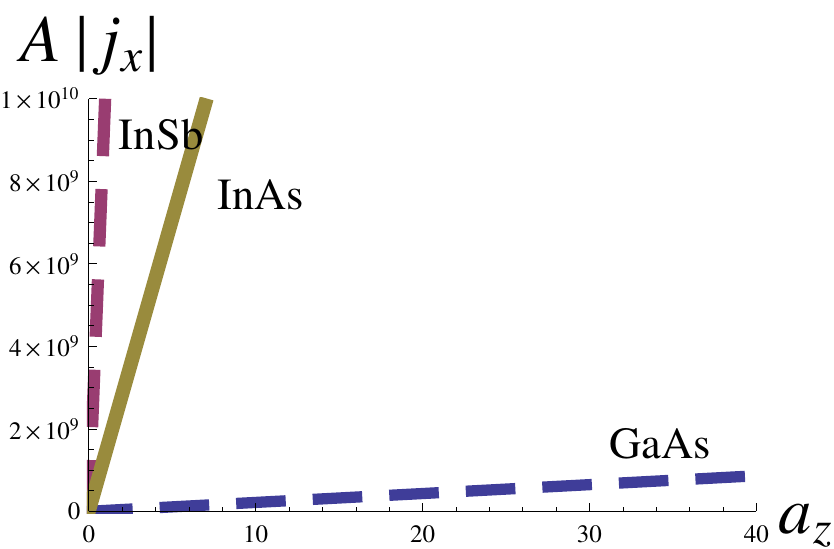}
\caption{\label{a} (Color online) Variation of spin current with acceleration for three different semiconductors, where A = $\frac{2m^{2}c^{2}}{\hbar e^{2}\tau^{2}\rho\mu}$.}
\end{figure}

Heisenberg's equation helps us to obtain the semi-classical force equation as
\beq \vec{F} = \frac{1}{i\hbar}\left[m^*\vec{\dot{r}},H \right] + m^*\frac{\partial\vec{\dot{r}}}{\partial t},\eeq with
$ \vec{\dot{r}}  = \frac{1}{i\hbar}[\vec{r}, H].$ Thus from (16)
\beq \vec{\dot{r}} = \frac{\vec{ p}}{m^*} -
\lambda_{eff}\frac{e}{\hbar}\left(\vec{\sigma}\times \vec{ E}_{tot}\right)\label{m}\eeq
Finally,
\begin{equation}\label{lor1}
\vec{F} = m^*\ddot{\vec{r}} = -e\vec{ \nabla}V_{tot}(\vec{ r})
+ \lambda_{eff}\frac{em^{*}}{\hbar}\dot{\vec{ r}}\times \vec{\nabla}\times
(\vec{\sigma}\times \vec{E}_{tot})
\end{equation}
is the spin Lorentz force with an effective magnetic field $\vec{\nabla}\times
(\vec{\sigma}\times \vec{E}_{tot}).$
Explicitly, the spin dependent vector potential is  given by
\begin{equation}\label{gauge}
\vec{A}(\vec{\sigma}) = \lambda_{eff}\frac{m^{*}c}{\hbar}(\vec{\sigma}\times \vec{E}_{tot}).
\end{equation}
The spin dependent effective Lorentz force noted in eqn.(\ref{lor1}) is responsible for the spin transport and the spin Hall effect of our inertial system. Obviously, the Lorentz force is enhanced due to $\vec{ k} . \vec{p }$ perturbation in comparison to the inertial spin force calculated in \cite{cb}.
The expression of $\dot{\vec{r}}$ in (18) can be rewritten as
\beq
\displaystyle {\dot{\vec{r}} = \frac{\vec{p}}{m} + \vec{v}_{\vec{ \sigma},~{\vec{a}}}} \eeq where
\begin{equation}\label{mnn}
\displaystyle \vec{v}_{\vec{\sigma},~\vec{a}}=
- \lambda_{eff}\frac{e}{\hbar}(\vec{\sigma}\times\vec{ E}_{tot})
\end{equation}
is the spin dependent anomalous velocity term, which depends on $\delta \lambda$ i.e on the spin orbit gap and the band gap energy of the crystal considered. The expression shows for a non zero spin orbit gap, the spin dependent velocity changes with the energy gap. For vanishing spin-orbit gap, there is no extra contribution to the anomalous velocity for the $\vec{ k} . \vec{ p }$ perturbation.
The spin current and spin Hall conductivity in an accelerated frame of a semiconductor can now be derived by taking resort to the method of averaging \cite{cb,n}. We proceed  with equation(\ref{lor1}) as $\vec{F} = \vec{F}_{0}+ \vec{F}_{\vec{\sigma}}$
where $ \vec{F}_{0}$ and $\vec{F}_{\vec{\sigma}}$ are respectively the  spin independent and the spin dependent parts of the total spin force.
The potential $V(\vec{r})$can be taken to be composed of the external electric potential $ V_{0}(\vec{r})$ and the lattice electric
potential $ V_{l}(\vec{r}) $.
If the momentum relaxation time $\tau$ is independent of $\vec{\sigma}$ and total electric field  $\vec{E}_{tot},$ is constant then for cubic symmetric semiconductor \cite{cb,n} i.e considering $\left\langle\frac{\partial^{2}V_{l}}{\partial r_{i}\partial r_{j}}\right\rangle = \mu \delta_{ij}\label{sm},$ we can write the current of the system
 with $\vec{k}.\vec{p}$ perturbation as
\beq \vec{j}_{kp} = \vec{j}^{o ,\vec{a}}_{kp} + \vec{j}_{kp}^{s,\vec{a}}(\vec{\sigma}) ,\eeq where
 \beq \vec{j}^{o , \vec{a}}_{kp} = \frac{e^{2}\tau \rho}{m^*}(\vec{E}_{0} - \vec{E}_{\vec{a}}) \label{jo} \eeq is the charge component of current and
\begin{eqnarray}\label{kol}
\vec{j}_{kp}^{s, \vec{a}}(\vec{\sigma}) & =& \frac{m}{m^*}
\left[1 + \frac{4m^{2}c^{2}P^{2}}{3\hbar^{2}}\left(\frac{1}{E_{G}^{2}} - \frac{1}{(E_{G} + \triangle_{0})^{2}}\right)\right]
\vec{j}^{s, \vec{a}}(\vec{\sigma})\nonumber\\
&=&  \frac{m}{m^*}(1 + \frac{\delta \lambda}{\lambda})\vec{j}^{s, \vec{a}}(\vec{\sigma})
\end{eqnarray}
is the spin current of this inertial system.
Here $\rho$ is the total charge concentration and $\vec{n}=\langle \vec{\sigma}\rangle$ is the spin polarization vector and
 $\vec{j}^{s, \vec{a}}(\vec{\sigma}) = \frac{\hbar e^{3}\tau^{2}\rho\mu}{2m^3c^2}\left(\vec{n}\times (\vec{E}_{0} - \vec{E}_{\vec{a}}\right)$ is the spin current in an  accelerating frame\cite{cb} without $\vec{k} . \vec{p}$ perturbation, where $\mu$ is a constant.
The ratio of spin current in an accelerating system with and without $\vec{k}.\vec{p}$ perturbation is given by
\beq \frac{|\vec{j}^{s, \vec{a}}(\vec{\sigma})_{kp}|}{|\vec{j}^{s, \vec{a}}(\vec{\sigma})|} = \frac{m}{m^*}(1 + \frac{\delta \lambda}{\lambda}).\eeq
The coupling constant $\delta \lambda$ is different for different materials and  $\lambda$, the coupling parameter in the vacuum  has a  constant value $3.7\times 10^{-6}{\AA}^{2}.$
Figure (\ref{a}) shows the variation of the ratio of spin current with acceleration for three different semiconductors.\\

The table reveals the fact that in a linearly accelerating frame, how $\vec{k} . \vec{p}$ method is useful for generating
large spin current in semiconductors.

Switching off the external electric field, we have
\beq \vec{j}_{kp}^{s, \vec{a}}(\vec{\sigma})  = - \sigma^{s,a}_{H, kp}
\left(\vec{n}\times  \vec{E}_{\vec{a}}\right), \eeq
where $\sigma^{s,a}_{H, kp} = \frac{2e^{3}\tau^{2}\rho\mu}{m^*\hbar}\lambda_{eff}$ is the spin Hall conductivity.
Considering the acceleration along $z$ direction, the spin current in the $x$ direction becomes \beq  |\vec{j}_{x,kp}^{s, \vec{a}}(\vec{\sigma})|  = \sigma^{s,a}_{H, kp}(n_{y}E_{a,z})\label{spin}\eeq

Eqn. (29) emphasize how the acceleration of the frame solely can produce huge spin current in a system.

It is clear that the spin conductivity is renormalized by the $\vec{k} .\vec{p}$ perturbation. The comparison of the spin Hall conductivity in our system with and without \cite{cb} $\vec {k}.\vec{p}$ perturbation  can be obtained as
\beq \frac{|\sigma^{s , \vec{a}}_{H},kp|}{|\sigma^{s, \vec{a}}_{H}|} = \frac{m}{m^{*}}(1 + \frac{\delta \lambda}{\lambda})\label{ratio}.\eeq This is an impressive result as it clearly indicate the importance of using semiconductors as spintronic materials. Besides, one can produce the large spin current through acceleration.

\section{Inertial Spin filter}
In the advancement of spintronic applications, a spin filter is an important candidate for producing spin polarized currents. Though there are lots of theoretical
attempts to describe a perfect spin filter in various aspects \cite{filter2,hatano}, but demonstration of  a perfect spin filter in an inertial system is new.
It is evident from the name, a spin filter is a spin based device to obtain polarized spins. We can write Hamiltonian (14) with an acceleration in the z direction and in presence of the external magnetic field $\vec{B}$ 
 as
\beq  H = \frac{\vec{\Pi}^{2}}{2m^*} + \frac{\alpha_{eff}}{\hbar}(\Pi_{x}\sigma_{y} - \Pi_{y}\sigma_{x}) ,\label{h} \eeq
where $\vec{\Pi} = \vec{p} - e \vec{A}(\vec{r}),$ and $\vec{B} = \nabla\times \vec{A}(\vec{r})$  and $\alpha_{eff} = \lambda_{eff}E_{a, z} $ is the Rashba \cite{rashba} like coupling parameter \cite{cb}, containing the effect of acceleration (in the z direction) as well as the $\vec{k}.\vec{p}$ parameters. The Hamiltonian in (\ref{h}) can also be written in the following form
\beq  H = \frac{1}{2m^{*}}\left(\vec{p} - \frac{e}{c}\vec{A}(\vec{r}) - \frac{q}{c}\vec{A}^{ ~'}(\vec{r}, \vec{\sigma})\right)^{2},\label{min}\eeq
 where
 \beq \vec{A}^{~ '}(\vec{r}, \sigma) = \frac{c}{2}(-\sigma_{y}, \sigma_{x}, 0) \label{moo},\eeq and  we have neglected the second order terms of $\vec{A}^{ ~'}(\vec{r}, \sigma).$
 $q = \frac{2m^*\alpha_{eff}}{\hbar}$ can be regarded as charge.
Obviously, the gauge due to external magnetic field provides an Abelian gauge but the spin gauge due to inertial SOC is non-Abelian in nature.
In terms of the total gauge field $\vec{\tilde{A}}^{~}(\vec{r})$, equation (\ref{h}) can be written as

\beq  H = \frac{1}{2m^{*}}\left(\vec{p} - \frac{\tilde{e}}{c}\vec{\tilde{A}}\right)^{2} \eeq
where $\vec{\tilde{A}} = e\vec{A}(\vec{r}) + q\vec{A}(\vec{r}, \vec{\sigma})$ and $\tilde{e}$ is a coupling constant, which is set to be $1$ for future convenience.
Using the total gauge $\vec{\tilde{A}}$  the Berry curvature can be written as
 \beq \Omega_{\lambda} = \Omega_{\mu \nu} = \partial_{\mu}\tilde{A}_{\nu} - \partial_{\nu}\tilde{A}_{\mu} -\frac{i\tilde{e}}{c\hbar}\left[\tilde{A}_{\mu}, \tilde{A}_{\nu}\right], \eeq which is similar to a physical field.
The $z$ component of this filed is
\beq \label{omega} \Omega_{z} =  \left(\partial_{x}\tilde{A}_{y} - \partial_{y}\tilde{A}_{x}\right) - \frac{i\tilde{e}}{c\hbar}\left[\tilde{A}_{x}, \tilde{A}_{y}\right].\eeq
$\Omega_{z} $ in our case boils down to the following form
\beq \Omega_{z} = eB_{z} + q^{2}\frac{c}{2\hbar}\sigma_{z} \label{abc}.\eeq
The second term in the expression of $\Omega_{z}$ in (\ref{abc}), actually represents a magnetic field in $z$ direction, with opposite sign for spin polarized along $+z$ direction or $-z$ direction.
As the spin up and spin down electrons experience equal but opposite vertical magnetic fields, they will subsequently carry equal and opposite $AC$ like phase \cite{casher} which can be obtained from \beq \phi_{AC} = \oint d\vec{r}. \vec{A}(\vec{r}, \sigma),\eeq  whereas the $AB$ phase appears due to the first term in (\ref{abc}) is the same for both up and down electrons.
The eqn (\ref{abc}) can be expressed in terms of the flux generated through area $W$ as
\beq \Omega_{z} = e\frac{\phi_{B}}{W} + q \frac{\phi_{I}}{W}, \label{abd}\eeq
where $\phi_{B} = WB_{z},$ flux due to external magnetic field and $\phi_{I} = Wq\frac{c}{2\hbar}\sigma_{z}$ is the physical field generated due to inertial SOC effect.
The first term on the right hand side(rhs) of (\ref{abd}) causes the $AB$ phase, whereas the second term on the rhs of (\ref{abd}) is the flux due to the physical field, is actually responsible for a $AC$ like phase. 

Thus the interplay of the AC like phase and AB phase can provide the explanation of inertial spin filter. In our framework, if we fix our AB phase as $\frac{\pi}{2}$ and the AC like phase as $\frac{\pi}{2}$ and $-\frac{\pi}{2}$ for up and down electrons, then the interference of up electrons is destructive in nature, whereas down electrons give the constructive interference. Finally, in the output we have down spin electrons only.


\section{Induced SOC via SRC}
In search of the answer to the question: are the different spin coupling terms interconnected, we start with the Hamiltonian (4) considering only the effect of rotation and in presence of an external magnetic field. We can then write
\beq H_{\Omega} = \frac{\vec{\pi}^{2}_{\Omega}}{2m}  - e\phi_{\Omega} - \mu\vec{\sigma} . \vec{B} - \frac{\hbar}{2}\vec{\sigma} . \vec{\Omega},\label{ome}\eeq
with $\vec{\pi}_{\Omega} = \vec{p} - e\vec{A}^{'} = \hbar \vec{k}_{\Omega},$ where $\vec{k}_{\Omega}$ is the modified crystal momentum and $\vec{A}^{'} = \vec{A} + \vec{A}_{\Omega},$ is the total gauge effective in the system.
It is well known that the rotation of the frame can induce a magnetic field \cite{barnett}. Here  $\vec{A}_{\Omega} = \frac{m}{e}(\vec{\Omega}\times \vec{r})$ corresponds to the rotation induced magnetic field $ \vec{B}_{\Omega} = \vec{\nabla}\times\vec{A}_{\Omega} = \frac{2m}{e}\vec{\Omega}$ and $\phi_{\Omega} = \frac{m}{2e}(\vec{\Omega}\times \vec{r})^{2}.$
The second term in the rhs of (\ref{ome}), is the potential term. The third term is the standard Zeeman coupling term induced due to the external magnetic field.
In terms of the rotation induced magnetic field, we can write the Hamiltonian in (\ref{ome}) as,
\beq H_{\Omega} = \frac{\vec{\pi}^{2}_{\Omega}}{2m} - e\phi_{\Omega} - \mu\vec{\sigma} . \vec{B} - \mu\vec{\sigma} . \frac{\vec{B}_{\Omega}}{2}. \label{d}\eeq This form of $\vec{B}_{\Omega}$ is similar to the gravitomagnetic field \cite{grav}, or the Barnett field given by $\vec{B}_{s} = \frac{m}{e}\vec{\Omega}$ \cite{matsuoprb}. The difference of these two fields lies in a factor of $2.$ The origin of the terms $\vec{A}_{\Omega}$ and $\phi_{\Omega}$ is the term $\vec{\Omega}.(\vec{r}\times \vec{p}),$ which is the coupling between  rotation and orbital angular momentum. The last term in Hamiltonian (\ref{d}) is the spin rotation coupling term which is effectively a Zeeman like term. Instead of standard external magnetic field, in this case the magnetic field is created due to the rotation of the system.

If we choose a planar rotation in the momentum space, we can still work with the same Hamiltonian as in (\ref{d}) and we get a $\vec{k}$ dependent Zeeman coupling. This $\vec{k}$ dependent Zeeman coupling is also experimentally verified quantity, which can be observed in real materials as $Sr_{2}RuO_{4}$ \cite{kdep}. This motivates us to study  the scenario of momentum dependent SRC term. The induced magnetic field can now be written as
\beq \vec{B}_{\Omega}(\vec{k}, t) = 2\frac{m}{e}\vec{\Omega}(\vec{k}_{\Omega}, t).\eeq
The Hamiltonian thus becomes
\beq H_{\Omega}(t) = \frac{(\vec{\pi}_{\Omega})^{2}}{2m} - e\phi_{\Omega}  - \mu\vec{\sigma} . \vec{B} -
 \frac{e\hbar}{4m}\vec{\sigma}.\vec{B}_{\vec{\Omega}}(\vec{k}_{\Omega},t)\label{ka}.\eeq
We can write the rotation frequency as $\Omega(\vec{k}_{\Omega}, t) = |\Omega(t)|\hat{n}(\vec{k}_{\Omega})$, where $|\Omega(t)|$ is the magnitude of rotation and $\hat{n}_{\Omega}(\vec{k}_{\Omega})$ is the unit vector along $\vec{k}_{\Omega}.$ Choosing $\hat{n}(\vec{k}_{\Omega}) = \frac{1}{|k|}(k_{x, \Omega},- k_{y, \Omega},0),$ the rotation induced magnetic field is given by \beq \vec{B}_{\Omega}(\vec{k}_{\Omega}, t) = 2\frac{m}{e}\frac{|\Omega|}{|k|}(k_{x, \Omega},- k_{y, \Omega},0),\label{e}\eeq where $|k| = \sqrt{(k_{x, \Omega})^{2} + (k_{y, \Omega})^{2}}.$
We can rewrite the Hamiltonian (45) as
\beq H_{\Omega} = \frac{(\vec{\pi}_{\Omega})^{2}}{2m}  - \mu_{B}\vec{\sigma} . \vec{B} - e\phi_{\Omega} +
\beta_{\Omega} ( \sigma_{y}\pi_{y,\Omega} - \sigma_{x}\pi_{x,\Omega})\label{sss}.\eeq
One may note that the last term in (\ref{sss}), represents a SRC induced SOC where the coupling constant 
$\beta_{\Omega} $ represents Dresselhaus like coupling parameter given by $\beta_{\Omega} = \frac{1}{2}\frac{|\Omega|}{|k|}$.
 %
Considering the effect of acceleration and rotation, the Hamiltonian can be written in the following form \cite{src,bc,cb}
\beq H = \frac{\hbar^{2}\vec{k}_{\Omega}^2}{2m} + \alpha_{a}(k_{x, \Omega}\sigma_{y} - k_{y, \Omega}\sigma_{x}) + \beta_{\Omega} (\sigma_{y}k_{y,\Omega} - \sigma_{x}k_{x,\Omega}) ,\label{as}\eeq
where the $\alpha_{a} = \lambda E_{a, z},$ is the Rashba like coupling parameter appearing due to acceleration as discussed in the previous section. In writing the above Hamiltonian there are two choices that one should opt, which are\\
i) The electric field due to acceleration should be in the $z$ direction.\\
ii) The rotation frequency must be momentum dependent and of specific form as mentioned above.\\

The knowledge of the relative strength of Rashba and Dresselhaus like terms is important for investigations of different spin dependent issues
in our inertial system. It is of interest that unlike the original Dresselhaus effect, we can externally control the $\beta_{\Omega}$ term. The k linear coupling terms give rise to the spin splitting scenario in semiconductors. So it is possible to control the spin splitting by controlling acceleration and rotation externally in our model.

\section{Inertial spin galvanic effect}
The spin galvanic effect(SGE) is due to the non-equilibrium but uniform population of electron spin no matter whether it is produced optically or by non-optical means. In this case it is possible to measure the anisotropic orientation of spins in the momentum space and hence the different
contribution of the Rashba and the Dresselhaus terms \cite{Ganichev1, Ganichev2}.
In spin Galvanic effect the necessary condition is the $k$ linear terms in the Hamiltonian. In our inertial system, the $k$ linear terms come through the acceleration and rotation terms. Thus the origin of the k linear terms are different. In semiconductor heterostucture these k linear terms arise because of the bulk inversion asymmetry(Dresselhaus term) and structural inversion asymmetry (Rashba term). We have Rashba like coupling from the acceleration of the system and Dresselhaus like term due to rotation.
As the
asymmetric spin levels are split, with one spin orientation favored, an excitation can
cause electrons to reverse spin, creating a charge current along the x-y plane.
From the Hamiltonian (\ref{sss}), one is able to calculate the total current as a function of the spin
polarization along the surface.
The $SGE$ current $j_{SGE}$ and
the average spin are related by a second rank pseudo-tensor
with components proportional to the parameters
of spin-orbit splitting parameter as follows \cite{Ganichev1, Ganichev2}
\begin{equation}\label{alf}
\vec{j}_{SGE} = A\left( \begin{array}{cc}
\beta_{\Omega} & -\alpha_{a} \\
  \alpha_{a} & \beta_{\Omega}
\end{array} \right)\vec{S}
\end{equation}
where $\vec{S}$ is the average spin in the plane and $A$ is a constant determined by the kinetics
of the $SGE$, namely by the characteristics of momentum
and spin relaxation processes. Unlike the usual cases, we have here two inertial spin orbit coupling strengths coming as a consequence of SOC terms due to acceleration and the spin rotation coupling effect. The spin galvanic current in (\ref{alf}) can be rewritten in terms of the acceleration and rotation parameters as
\begin{equation}\label{al}
\vec{j}_{SGE} = A\left( \begin{array}{cc}
\frac{|\Omega|}{2|k|} & - \frac{\hbar^{2}}{4mec^2}a_{z} \\
  \frac{\hbar^{2}}{4mec^2}a_{z} & \frac{|\Omega|}{2|k|}
\end{array} \right)\vec{S}
\end{equation}
Eqn. (\ref{al}) helps one to find out the ratio of $\alpha_{a}$ and $\beta_{\Omega}$ or the ratio of acceleration and rotation. The spin galvanic current can be decomposed into the current duo to Rashba like term and Dresselhaus like term $\vec{j}_{a}$ and $\vec{j}_{\Omega}$ proportional to the $\alpha_{a}$ and $\beta_{\Omega}$ term. From symmetry, $\vec{j}_{a}$ is perpendicular to average spin $\vec{S}$, whereas $\vec{j}_{\Omega}$  makes an angle $2\Psi$ with $\vec{S}$, where $\Psi$ is the angle between $\vec{S}$ and the $x$ axis.
Thus one can find out the spin galvanic currents in terms of the coupling terms due to acceleration and rotation. The absolute value of the total current $j_{SGE}$
is given by the expression
\beq j_{SGE} = \sqrt{j_{a}^{2} + j_{\Omega}^{2} - 2j_{a}j_{\Omega}sin 2\Psi}\label{mq}\eeq
Now suppose the average spin vector is oriented along the $x$ axis, then we have angle $\Psi = 0$. Thus from (\ref{mq}), we have $\vec{j}_{a}$ and $\vec{j}_{\Omega}$ in the directions perpendicular and parallel to $\vec{S}$, respectively. The ratio of the currents in the $x$ and $y$ direction is given by
\beq \frac{j_{y}}{j_{x}} = \frac{\alpha_{a}}{\beta_{\Omega}} ,\eeq
or interestingly
\beq \frac{a_{z}}{|\Omega|} \propto \frac{j_{y}}{j_{x}}.\eeq
The ratio of acceleration to rotation can be achieved from the above equation. Thus we can control the charge current flowing through the system through the two non-inertial parameters. On the other hand, from the observed ratio of the currents one can understand whether acceleration or rotation contribution is dominating in the system. We should mention here that for different orientation of $\vec{S}$ we get different values of this ratio. It may be noted that, without any external electromagnetic fields we are successful to produce charge current from the $k$ linear terms of the Hamiltonian, due to inertial effects.

\section{Spin relaxation time}
The electron spin relaxation mechanism is based on spin-orbit-coupling. In this section we want to investigate how the inertial SOC terms and the SRC induced SOC term affect the spin relaxation procedure. In particular, our motivation is to analyze the role of Rashba like coupling $\alpha_{a}$ and Dresselhaus like coupling parameter $\beta_{\Omega}$ in the expression of spin relaxation time. Let us now consider a different type of $k$ dependence of the rotation frequency as $\Omega(\vec{k}_{\Omega}, t) = |\Omega(t)|(k_{x, \Omega}k_{y, \Omega}^{2},- k_{y, \Omega}k_{x, \Omega}^{2},0)$ with which we can write the rotation induced magnetic field as
\beq \vec{B}_{\Omega}(\vec{k}_{\Omega}, t) = 2\frac{m}{e}\frac{|\Omega|}{|k|}(k_{x, \Omega}k_{y, \Omega}^{2},- k_{y, \Omega}k_{x, \Omega}^{2},0).\label{ef}\eeq With this magnetic field  and considering the $k$ linear coupling term due to acceleration we can write the Hamiltonian (\ref{as}) as \beq H = \frac{\hbar^{2}\vec{k}_{\Omega}^2}{2m} + \alpha_{a}(k_{x, \Omega}\sigma_{y} - k_{y, \Omega}\sigma_{x}) + \beta_{\Omega} (\sigma_{y}k_{y, \Omega}k_{x, \Omega}^{2} - \sigma_{x}k_{x, \Omega}k_{y, \Omega}^{2}) .\label{asf}\eeq The last term reminds us about the $k^{3}$ Dresselhaus like coupling but with a different origin.
Now if we consider a spin-independent impurity potential as $V_{i}(\vec{r})$ in the Hamiltonian then we can write
\beq H = \frac{\hbar^{2}\vec{k}_{\Omega}^2}{2m} + \alpha_{a}(k_{x, \Omega}\sigma_{y} - k_{y, \Omega}\sigma_{x}) + \beta_{\Omega} (\sigma_{y}k_{y, \Omega}k_{x, \Omega}^{2} - \sigma_{x}k_{x, \Omega}k_{y, \Omega}^{2}) + V_{i}(\vec{r}) .\label{af}\eeq
For a disordered and non-interacting SOC system the spin dynamics is elucidated through the spin diffusion equations. These equations can be derived in the density matrix formalism. In this formalism at initial time t=0, we can write the non-equilibrium spin density distribution by considering the length scale much larger than the mean free path as
\beq S_{m}(r,t) = \int d\vec{r}^{'} P_{mn}(t, r,\vec{r}^{'})S_{n}(0,\vec{r}^{'}),\label{s}\eeq
where $S_{n}(0,\vec{r}^{'})$ is the initial spin density at $\vec{r}^{'}$ and the diffusion Greens function is denoted by $P_{mn}(t, r,\vec{r}^{'}).$ In the momentum space and for homogeneous system equation (\ref{s}) can be written as
\beq [I - P(\omega, \vec{k})].\vec{S} = 0,\eeq where $I$ is the unit matrix and $P(\omega, \vec{k})$ is the Fourier transformed diffusion Greens function. In a non-inertial frame in the low frequency limit the sets of spin diffusion equations for time and space dependent spin densities are given by \cite{cin}
\begin{eqnarray}
\frac{\partial S_{x}}{\partial t} &=& \frac{v_{F}^{2}}{2}\nabla^{2}S_{x} + \frac{8v_{F}^{4}m\tau}{k_{F}\alpha_{a} \beta_{\Omega}}\frac{1}{\left[\frac{k_{F}^{2}}{\alpha_{a}^{2}} + \frac{32}{k_{F}^{2}\beta_{\Omega}^{2}}\right]}S_{y} +\left(\frac{2mv_{F}^{4}\tau}{\alpha_{a}}\frac{\partial}{\partial x} + \frac{8mv_{F}^{4}\tau}{k_{F}^{2}\beta_{\Omega}}\frac{\partial}{\partial y}\right)S_{z} - \frac{S_{x}}{\tau_{s}}\nonumber\\
\frac{\partial S_{y}}{\partial t} &=& \frac{v_{F}^{2}}{2}\nabla^{2}S_{y} + \frac{8v_{F}^{4}m\tau}{k_{F}\alpha_{a} \beta_{\Omega}}\frac{1}{\left[\frac{k_{F}^{2}}{\alpha_{a}^{2}} + \frac{32}{k_{F}^{2}\beta_{\Omega}^{2}}\right]}S_{x} + \left(\frac{2mv_{F}^{4}\tau}{\alpha_{a}}\frac{\partial}{\partial y} + \frac{8mv_{F}^{4}\tau}{k_{F}^{2}\beta_{\Omega}}\frac{\partial}{\partial x}\right)S_{z} - \frac{S_{y}}{\tau_{s}} \nonumber\\
\frac{\partial S_{z}}{\partial t} &=& \frac{v_{F}^{2}}{2}\nabla^{2}S_{z} - \left(\frac{2mv_{F}^{4}\tau}{\alpha_{a}}\frac{\partial}{\partial x} + \frac{8mv_{F}^{4}\tau}{k_{F}^{2}\beta_{\Omega}}\frac{\partial}{\partial y}\right)S_{x}\nonumber\\
&& ~~~~~~~~~~~~~~~~~~~~~~~~~~~~~~~-\left(\frac{2mv_{F}^{4}\tau}{\alpha_{a}}\frac{\partial}{\partial y} + \frac{8mv_{F}^{4}\tau}{k_{F}^{2}\beta_{\Omega}}\frac{\partial}{\partial x}\right)S_{y} - \frac{2S_{z}}{\tau_{s}},\label{ss}
\end{eqnarray}
where $v_{F}$ is the Fermi velocity, $\tau$ is the momentum relaxation time and $\tau_{s}$ is the spin relaxation time given by
\beq \tau_{s} = \frac{1}{2v_{F}^{4}\tau}\frac{1}{\left[\frac{k_{F}^{2}}{\alpha_{a}^{2}} + \frac{32}{k_{F}^{2}\beta_{\Omega}^{2}}\right]}\label{23} .\eeq As it is evident from (\ref{23}), the spin relaxation time $\tau_{s}$ depends on the coupling terms $\alpha_{a}$ and $\beta_{\Omega},$ which actually represents the non-inertial parameters. Tuning of the non-inertial parameters may help us to obtain larger spin relaxation time, which is of utmost importance.

 \section{Conclusion}
In this paper we have discussed the generation of spin current from the covariant Dirac equation in a linearly accelerating system in the presence of electromagnetic fields with the $\vec{k}. \vec{p}$ theory. Interestingly, the acceleration of the frame can induce a SOC term apart from the SOC term induced by the external electric field. This induced SOC is enhanced due to the $\vec{k}. \vec{p}$ perturbation. The expressions for the inertial spin current and conductivity are derived. The effect of acceleration is explicit in the expressions for the spin component of current. For a constant external electric field, the spin current is found to vanish for a particular value of acceleration $a_c.$ Furthermore, from the gauge theoretical point of view, discussion of a perfect spin filter in an inertial system is made. In the context of inertial effects on spin transport, we have discussed the generation of a new type of SOC coupling induced from momentum dependent SRC term. Taking resort to this SRC induced SOC term, which is of Dresselhaus type, we can successfully discuss the inertial spin Galvanic effect in our system. Besides, the effect of inertial parameters in the control of spin relaxation time is also described. It is hoped that the manipulation of spin current and related effects through the non-inertial parameters may trigger up various future applications in spintronic devices.

\newpage
\begin{tabular}{|*{2}{c|}l|l|l|}
\hline
$E_{G}(eV)$ & $\triangle_{0}(eV)$ & $P(eV\AA)$ & $\delta \lambda (\AA^{2})$ & $\frac{|\vec{j}^{s, \vec{a}}(\vec{\sigma})_{kp}|}{|\vec{j}^{s, \vec{a}}(\vec{\sigma})|}$ \\
\hline
GaAs = 1.519 & 0.341 & 10.493 & 5.3&$   2.154\times10^{7} $ \\ \hline
AlAs = 3.13  & 0.300 & 8.97  &0.318 & $  5.748\times10^{5} $\\ \hline
InSb = 0.237 & 0.810 & 9.641 &523.33 &$   1.0175\times10^{10} $\\\hline
InAs = 0.418 & 0.380 & 9.197 &120 &$  1.41\times10^{9} $ \\
\hline

\end{tabular}\\
\begin{center}
\textbf{Table:} Table of different Kane model parameters and ratio of current with and without $\vec{k}.\vec{p}$ perturbation for different semiconductor
 \end{center}
\end{document}